\documentclass{article}    

\usepackage{graphicx}

\title{\textbf{Quantum Localization}}  
\author{Richard A. Mould\footnote{Department of Physics and Astronomy, State University of New York, Stony Brook,
\mbox{New York} 11794-3800; http://ms.cc.sunysb.edu/\~{}rmould}}  
\date{}    
   
\begin{document}             

\maketitle              

\begin{abstract}

The auxiliary q-rules of quantum mechanics developed in other papers are applied to the problem of the location of material
objects \mbox{-- both} macroscopic and microscopic.  All objects tend to expand in space due to the uncertainty in their momentum. 
The q-rules are found to oppose this perpetual expansion with a collapse mechanism that insures the dependable localization of  objects in
ordinary human experience.

\end{abstract}

\section*{Introduction}
Object localization is essential if macroscopic objects are to be limited in space.  This property does not follow from the Schr\"{o}dinger equation by itself, for objects that are subject only to that equation will expand forever due to their uncertainty in momentum.  This may seem to be no problem in a universe consisting of objects that are already well localized.  But our universe evolved from one that was not well localized.  Go back to the time of recombination some 300 thousand years after the big bang when the only atomic structures are hydrogen and helium.  These atoms must have had a very large uncertainty of position at that time; and under the influence of Schr\"{o}dinger's equation, subsequent evolution would have spread them out even further.  The Schr\"{o}dinger equation supports correlations but not contractions of the wave function, so it cannot reduce an existing uncertainty of position.  For this reason it could not have produced the well-localized macroscopic objects that populate the contemporary universe.  Therefore, the dynamic principle of quantum mechanics \emph{must include, or must be supplemented by a collapse mechanism}.

\pagebreak

\section*{A box of molecules}
	Imagine a gravity-free box that initially contains only $N$ molecules of hydrogen and $N$ molecules of chlorine, where the position of each molecule is entirely uncertain.  The wave function of each molecule fills the box completely and uniformly.  In time, separate molecules will disappear and be replaced by hydrogen chlorine gas molecules, where the wave function of each molecule also fills the box completely and uniformly. The Schr\"{o}dinger equation establishes the correlations that bring the hydrogen and the chlorine atoms together, but it is incapable of localizing the result to a volume that is less than the uncertainty of the initial ingredients.  

Take this reasoning a step further.  Imagine a gravity-free box that contains all the atoms  and molecules that are necessary to make a macroscopic 1 gm rock, where each of these ingredients initially exists independent of the others.  Let the initial position of each atom or molecule be entirely uncertain so the wave function of each fills the box completely and uniformly.  After a long period of time and many collisions within the box, the Schr\"{o}dinger equation will establish the necessary correlations to make a fully formed 1 gm macroscopic rock.  The box would then contain no separate molecules or atoms -- only a single rock.  The question is: Where would that rock be located inside the box?  Following the above example we would have to conclude that the rock, like all of its initial ingredients, would fill the box completely and uniformly.  It would be a macroscopic object that has a completely uncertain quantum mechanical position inside the box.

This is an unbelievable conclusion.  It is not illogical given the dynamics of quantum mechanics, but it certainly seems unphysical.  When that logic is applied to the entire universe it gives the results alluded to in the first paragraph.  The contemporary universe would then contain the standard variety of macroscopic objects including stars, planets, asteroids, and galaxies; but the location of each would be wildly uncertain.  

The Copenhagen view of quantum mechanics proposed by Niels Bohr and others recognizes the collapse of quantum mechanical systems in individual cases, and attributes this to their interaction with macroscopic scientific instruments.  But that cannot be the only mechanism of collapse; because in that case, there would be no way that a collapse process could get started in a universe like ours -- with no macroscopic objects to begin with.  

\pagebreak

	My claim is that Schr\"{o}dinger's equation \emph{can support a macroscopic location superposition just as easily as it can support a microscopic location superposition}, and it will do so unless there is a collapse mechanism that opposes it.  Therefore, the only reason we don't now experience rock superpositions left over from the time of recombination is that contemporary rocks must have undergone a great many wave contractions over the eons.  These contractions must have occurred without the help of pre-exiting localized macroscopic objects, for such objects did not exist 300 thousand years after the big bang.  Foundation theory must either alter the Schr\"{o}dinger equation or add auxiliary rules that provide for a collapse.  

	This requirement is satisfied by the `spontaneous localization' theory of Ghirardi and Pearle that is perhaps the most accepted alternative to the Copenhagen idea about collapse \cite{GPR}.  According to the theory, every atom in a macroscopic object undergoes a continuous process of spontaneous localizations; and because of correlations that it has with the object's other atoms, the entire object is continuously localized.  The stochastic process that produces this result is built directly into the Ghirardi-Pearle Hamiltonian of the system, so it is an integral part of the dynamic principle.  This theory is probably better understood in terms of the original model that relies on discontinuous localizations or `hits' to the system \cite{GRW}.  Accordingly, each atom of the object experiences a stochastic hit that causes it to spontaneously localize.  This happened to a single atom only once in every $10^{16}$ seconds.  But as a result, a macroscopic object experiences as many as $10^7$ hits per second, and each time that happens the entire object is localized because of position correlations between the gross object and each of its atoms.  These ideas have some unique experimental consequences that have not yet been verified \cite{RM}.

\section*{Q-rule localization}
	The q-rules also provide for frequent state reductions that affect microscopic objects and (through correlations) macroscopic objects as well.  Instead of showing this for a macroscopic 1 gm. rock, the point is illustrate by showing how the q-rules affect state reductions of free atoms like the original hydrogen or helium atoms that occupied the early universe.  

Let a photon raise an atom (primordial or contemporary) to an excited state, after which the atom drops down again by spontaneously emitting a photon.  The atom is assumed to be spread out widely over space by an amount that exceeds its minimum volume.  This is defined to be the smallest volume that the atom can occupy consistent with its uncertainty of momentum.  The atom in Fig.\ 1 (shaded area) is assumed to have spread far beyond this volume before it interacts with the photon.  As the incoming photon passes over the enlarged atom, scattered radiation will appear as a result of an interaction.  This will result in a superposition of many photons that originate from different parts of the atom's extended volume as shown in Fig.\ 1 -- these are the small wavelets in the figure.  The original correlations between the nucleus of the atom and its orbiting electrons must be preserved, even though the atom is spread out over a much larger volume.  That is, the smaller dimensions of the minimal volume atom must be unchanged during its expansion, so the potential energy of the orbiting electrons is unchanged.  The atom could not otherwise act as the center of a `characteristic' photon emission.  This means that the incident photon will engage the \emph{compact} atom throughout every part of the enlarged volume.

\begin{figure}[h]
\centering
\includegraphics[scale=0.8]{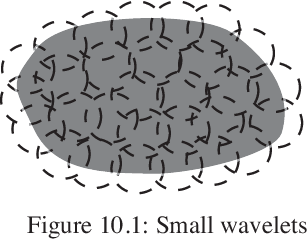}
\center{Figure 1:}
\end{figure}

Equation 12 in another paper  describes how an atom $a$ will respond to a laser field of $N$ photons \cite{RM1}. 
  
\begin{displaymath}
\Psi(t \ge  t_0) = \gamma_Na_0 \Leftrightarrow \gamma_{N - 1}a_1 + \gamma_{N - 1}\underline{a}_0 \otimes \gamma
\end{displaymath}
where $a_0$ is the atom in its ground state and $a_1$ is the atom in its excited state.  The state $\gamma_n$ is the field of $N$ gammas. The first and second components represent the stimulated oscillation between the two levels, and the third component is the spontaneous emission of a photon $\gamma$.

The same will be true of the atom in Fig.\ 1, except that the spontaneous emission part of that equation will be the sum of the probabilities of emissions coming from different parts of the extended atom.  The q-rule equation for the total process is therefore given by

\begin{displaymath}
\Psi(t \ge t_0) = \gamma_Na_{0} \Leftrightarrow \gamma_{N - 1}a_{1} + \lim_{n\to\infty}\Sigma_n\gamma_{N -
1}\underline{a}_{0n}\otimes\gamma_n
\end{displaymath}
where $\sum_n$ is a sum over all the ways that the atom can spontaneously emit a photon (i.e., all the wavelets in the figure), and $a_{0n}$ refers to each associated minimum volume atom.  As the sum over $n$ goes to infinity, the probability current flowing into each term in the summation goes to zero in such a way as to preserve the total square modulus.

The wavelets in Fig.\ 1 are represented by ready components in the above equation, so they are physically unreal.  If a stochastic hit occurs at time $t_{sc}$ on the minimum volume atom $a_{0k}$, the equation becomes
\begin{displaymath}
\Psi(t \ge t_{sc} > t_0) = \gamma_{N-1} a_{0k}\otimes\gamma_{k}
\end{displaymath}

So the atom is \emph{reduced to its minimum volume} in this interaction.  This may be different from the `initial' minimum volume because $\Delta \mathbf{p}$ of the atom might have changed during its interaction with the radiation field.

It is important that the q-rules provide a mechanism for a reduction of this kind without introducing artificial notions such as previously localized macroscopic instruments.  The rules provide an automatic contraction mechanism to counteract the automatic expansion mechanism of Schr\"{o}dinger's equation.

The q-rules also predict a unique experimental result that has not yet been verified and that is in direct conflict with the predicted Ghirardi-Pearls results.  This experiment and a comparison with the G-P predictions is outlined in another paper (Ref. 3).

\end{document}